\documentclass[aip,jcp,preprint]{revtex4-1}   
 
\usepackage{graphicx}
\usepackage{dcolumn}
\usepackage{bm}
\usepackage{color}
\usepackage{graphics,epsfig}
\usepackage{amsmath}
\usepackage{multirow}                              
\usepackage{setspace} 	                        
\usepackage{epsfig}
\usepackage{xspace}
\usepackage[export]{adjustbox}
\usepackage{epstopdf}
\usepackage[xspace=true]{chemmacros}
\usepackage{cleveref} 

\input colordvi.tex
\graphicspath{{./}{figures/}}

\newcommand*{\citen}[1]{%
  \begingroup
    \romannumeral-`\x 
    \setcitestyle{numbers}%
    \cite{#1}%
  \endgroup   
}
\begin{document}

\setcounter{page}{1}

\title{\vspace{-15mm}\fontsize{18pt}{10pt}\selectfont\textbf{Convolutional neural network approach to ion Coulomb crystal image analysis}} 

\date{\today}
\author{ James Allsopp$^2$, Jake Diprose$^3$, Brianna R. Heazlewood $^3$, Chase Zagorec-Marks$^{4,5}$, H.~J. Lewandowski$^{4,5}$, Lorenzo S. Petralia$^6$     Timothy P. Softley}
\thanks{Corresponding author. Email: t.p.softley@bham.ac.uk}
\affiliation{
$^1${School of Chemistry, University of Birmingham, Edgbaston, B15 2TT, United Kingdom};\\
$^2${Research Software Group, University of Birmingham, Edgbaston, B15 2TT, United Kingdom};\\
$^3${Department of Physics, University of Liverpool, Liverpool, L69 7ZE, United Kingdom};\\
$^4${JILA, National Institute of Standards and Technology and the University of Colorado, Boulder, CO 80309, USA};\\
$^5${Department of Physics, University of Colorado, Boulder, Colorado 80309, USA};\\
$^6${Institute for Breath Research, University of Innsbruck, Innsbruck, Austria};\\}

\begin{abstract}
{\bf Abstract}\\
This paper reports on the use of a convolutional neural network (CNN) methodology to analyse fluorescence images of calcium-ion Coulomb crystals in the gas phase. A transfer-learning approach is adopted using the publicly available RESNET50 model. It is demonstrated that by training the neural network on around 500,000 simulated images, we are able to determine ion-numbers not only for a verification set of 100,000 simulated images, but also for experimental calcium-ion images from two different laboratories using a wide range of ion-trap parameters.  Absolute ion numbers in the crystal were determined for the experimental data with a percentage error of approximately 10\%.   This analysis can be performed in a few seconds for an individual crystal image, and therefore the method enables the objective, and efficient, analysis of such images in real time, thereby facilitating time-dependent kinetic measurements on ion-molecule chemistry. 
The approach adopted also shows promising performance for identifying Ca$^+$ ion numbers in images of mixed-species crystals. 
\end{abstract}

\maketitle


\section{Introduction}\label{sec:intro}
\subsection{Formation and imaging of laser-cooled ionic Coulomb Crystals}\label{subsec:coulomb_crystals}
The uses of deep-learning methods to aid analysis of experimental results in the physical sciences have been growing rapidly in recent years.  
This is particularly true for image analysis \cite{Zhu2024,Ziatdinov2022,Sokolov2024}, where very large data sets may be generated, and there is a need to extract parameters that characterise the data efficiently and rapidly, even in real time as the data is being recorded.
In the current paper, we report the use of a deep-learning method with a convolutional neutral network (CNN) to analyse fluorescence images of ion Coulomb crystals (CC),  which are created when ions in the gas phase are laser-cooled in a multipole radiofrequency trap (such as a Paul trap \cite{paul1990a}). We demonstrate that the efficiency and objectivity of analysis of the images - in terms of deriving the absolute numbers of ions of various chemical species in a given crystal -  can potentially be dramatically enhanced through the use of a  CNN. We also show that a publicly available CNN model, RESNET50, which was originally trained to categorise images of everyday objects, animals, complex scenes etc., can be retrained to analyse and extract numerical data from scientific images of this type. 

\begin{figure}[hbt!]
    \centering
    \includegraphics[width=0.8\textwidth]{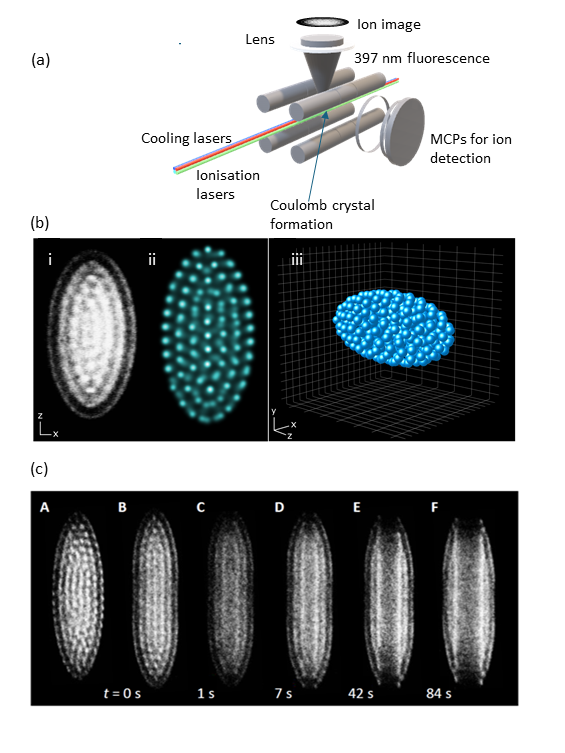}
    \caption{Coulomb crystal image generation and simulation.  (a) Ion trap with imaging set up (calcium oven and effusive beam not shown). (b)(i) Experimental Ca$^+$ ion fluorescence image of Coulomb crystal. (ii) Matching simulated image using the Molecular Dynamics code CCMD (see text for details). (iii) A simulated 3D representation of the Coulomb crystal, from the CCMD code, corresponding to the image shown in (ii).  (c) A sequence of experimental images: (A) Pure Ca$^+$ crystal; (B) 'flattened' crystal with Xe$^+$ sympathetically cooled on the outside; (C-F) progressive time sequence of images as Xe$^+$ reacts with NH$_3$ to form NH$_3^+$ illustrating the growing hollow core of non-flurorescing NH$_3^+$.  \cite{petralia2020} }
    \label{fig:CC_intro}
\end{figure}

As discussed previously in several  articles \cite{drewsen2003,willitsch2012,heazlewood2015,zhang2017,lewandowski2021}, ionic Coulomb crystals are normally formed within a quadrupole (Paul) trap, with the ions created in the trap  by photoionization of neutral precursor atoms or molecules, at low densities in an ultrahigh vacuum (\Cref{fig:CC_intro}(a)). The trap applies confining forces in three dimensions with a combination of radiofrequency and electrostatic fields,  tending to push the ions together, but the compression of the ion cloud is opposed by the repulsion between like-charged ions.
Initially on formation, the packet of ions forms a diffuse, disordered, and mobile cloud because the kinetic energy of the ions exceeds any localised potential energy minima.
However, when laser cooling is applied to the ions, most commonly to the alkaline-earth singly-charged ions (Ca$^+$ in the work described here), the temperature can be lowered,  typically to the milliKelvin range, where a phase transition occurs
to a  pseudo-crystalline structure. The ions take up sites on an ordered 3D ellipsoidal multilayered array ~\textendash~  a gas phase ‘crystal’ ~\textendash~ typically with just a few hundred ions and an ion-to-ion spacing of order 10~\textendash~20~$\mu$m (cf the spacing in a true solid-state crystal, which is of order 0.2~nm). The phase change occurs so as to minimise the potential energy of the system through the ions taking up  regular lattice positions at the optimum separation.
\Cref{fig:CC_intro}(b)(iii) shows a simulated 3D structure of  a Coulomb
crystal of Ca$^+$ ions of approximate total length 300~$\mu$m on the long axis. 

The laser-cooled ions are constantly fluorescing through the optical excitation and de-excitation cycle, and the rate of photon emission and the large (micron-scale) separation of the ions is sufficient to enable 
a simple microscope with an imaging camera to be used to observe layers of the crystal structure with single-ion resolution.
\Cref{fig:CC_intro}(b) (i) shows an example experimentally recorded  Ca$^+$ ion image  in which each white, diffuse spot represents a single fluorescing Ca$^+$ ion that is located in the central 2D slice
through the CC in the imaging-system focus.
 Various applications of ionic Coulomb crystals have been developed in spectroscopy, reaction dynamics, and kinetics \cite{drewsen2003,willitsch2012,heazlewood2015,zhang2017,lewandowski2021,Willitsch2024}, and the study of cold ion-molecule collisions has become an important component of the burgeoning field of cold and ultracold chemistry \cite{toscano2020,Softley2024,heazlewood2021}. Such experiments take advantage of the fact that
other types of positively charged ions, beyond those that can be laser-cooled, can be sympathetically cooled and trapped in the
crystal (e.g., NH$_3^+$, C$_2{\rm H}_2^+$, OCS$^+$,
 HN$_2^+$, MgH$^+$). In general, these other ions are not directly visible in the images
because they are not fluorescing under the experimental conditions. However, they tend to either
be found as a dark inner shell of the crystal (if they are lighter mass than the laser-cooled Ca$^+$) or as a dark outer shell (if heavier). Hence, their presence can be detected by the observed deformation of the fluorescing Ca$^+$ ion framework.
Cold ion-molecule reactions can be studied by passing cold, reacting neutral gases through the CC \cite{willitsch2012,gingell2010,Rosch2016,Kocheril2025}, or superimposing the ion trap with a trap for neutral atoms or molecules \cite{Willitsch2024}, and observing how the images change with time.
\Cref{fig:CC_intro}(c) shows a sequence of images that was recorded over two minutes, in which the initial pure Ca$^+$
crystal (image A) is ‘doped’ with Xe$^+$ ions that sit around the outside (note the flattened Ca$^+ $ structure
in images B and C). The Xe$^+$ ions then react with NH$_3$ (ammonia) by charge transfer to form NH$_3^+$ and a ‘hollow core’ appears in the
right-hand images (D-F), because the NH$_3^+$
 products of the charge transfer are trapped in the centre
of the crystal. \cite{petralia2020,tsikritea2021}

The extraction of a reaction rate from such measurements  may be achieved for relatively simple reactions (particularly those with only one possible product ion) by recording 
numerous images of the type shown in \Cref{fig:CC_intro}(c) as a function of reaction time. The 2D images can then be analysed to determine the number of ions of each  chemical species (mass)
 in the full 3-dimensional CC structure, at a sequence of  time points.
One attractive feature of this methodology is that it is a non-destructive detection method, and thus, in principle, a reaction rate can be derived from measurements on a single Coulomb crystal \cite{willitsch2008cold,gingell2010,bell2009}.
 \subsection{Determining ion numbers of Coulomb crystals}
There are a number of reasons why the calcium-ion numbers can not be routinely read directly from the CC images (unless the crystal contains only a very small number of ions) or determined accurately by simply measuring the overall geometry of the crystal image.
\begin{itemize}
\item{The images are a two-dimensional cut in the image plane through the 3-dimensional ellipsoidal crystal, and do not fully show the detailed 3D structure. Blurred features lying outside the focal plane appear in the crystal images to an extent that depends on the depth of focus in the imaging.}
\item{As the  number of ions is increased,  the crystal dimensions do not change monotonically with ion number but rather undergo step changes in the dimensions when the number of layers increases. This is particularly the case when there are just a few layers in the crystal. The ion numbers at which these step changes occur vary with the applied fields, trap geometry, and temperature.}
\item {The crystal structure has a dependence on the applied RF field and the endcap voltages, which increasingly squeeze the crystal in the $x-y$ or $z$ dimensions respectively as the RF peak-to-peak voltage or endcap voltages are increased. Thus, the number of ions shown in the 2D images changes as these voltages are adjusted, as well as the size and shape of the overall crystal image}.
\item{In practice, the effective temperature of the ions in the crystal may change experimentally due to a number of possible effects including background collisions with residual gases in the vacuum, and imperfect overlap of laser beams. This may lead to blurring of the images, and difficulty identifying all the spots in an image.}
\item {The structure is also subject to experimental imperfections due to patch charges on the electrodes or unsymmetrical positioning of the crystal with respect to the trap centre, or even due to imperfections on the imaging detector, leading to distortions of the images.}
\end{itemize}

To analyse the images for the number of ions,  the crystals can be simulated using an appropriate molecular dynamics code, which produces simulated images for given ion trap parameters and ion numbers.  \Cref{fig:CC_intro}(b)(ii) shows the result from a molecular dynamics (MD) simulation of a pure Ca$^+$ crystal image. Such a simulated image can then be juxtaposed with the experimental image, and  the number of ions determined by iteration of the initial parameters for the simulation to get a good match. We will refer to this hereafter as the method of {\it iterative simulated-image matching}. Examples of the end result of this matching can be seen later in \Cref{fig:Inference_for_Boulder}. In general,  the matching of experimental and simulated images is performed by the human eye, rather than an automated procedure. 
In practice, this is an extremely time-consuming method of analysis - for crystals with a few hundred ions, each MD simulation might take several CPU hours to generate a single simulated image and the simulation time scales up in proportion to $N^2$ (where $N$ is the number of ions); and then this comparison of experiment and simulation needs to be iterated.  Furthermore, there is inevitably a degree of subjectivity in matching by eye a `perfect' simulated image with an `imperfect' experimental one, especially when trying to define the ion numbers with a percentage error of better than  5\%.

An alternative, complementary method for determining ion numbers in the crystal was developed \cite{meyer2015,Schmid2017,Schneider2014,Rosch2016}, in which the ions are near-instantaneously ejected from the trap into a time-of-flight mass spectrometer and the number of ions arriving at the detector are measured in mass-selective time gates.  
While such measurements can be made in real time and indicate directly the mass and relative proportions of all species that are present \cite{Zagorec-Marks2024,Kocheril2025}, one disadvantage of this approach is that the CC is destroyed in the process; therefore this can  be used to determine ion numbers at only a single time point in a reaction kinetics experiment. To make measurements at a number of time points, new crystals must be formed for each time point, and, in practice, there will be variations in the initial numbers of ions in each of these crystals that needs to be accounted for to determine the reaction kinetics.
In addition, the accuracy of this method relies on calibration of the ion numbers at the detector (and the extraction efficiency) and, over a long period of time, the absolute accuracy of ion numbers may change due to degradation of the detector, even if the relative ion-numbers for different masses remain accurate. The detection efficiency of ions of different masses may also differ, especially if a wide spread of ion masses is present \cite{Zagorec-Marks2024,diprose2024}.

Ideally, one would like to have a real-time, non-destructive method of analysis of an experimental sequence of images.  
In this paper, we report on exploring the use of a deep-learning approach using a CNN to enable rapid and reproducible analysis of ion numbers from Coulomb-Crystal images. This would bring a greater level of objectivity to the process of extracting ion numbers from an image. It might also allow rapid deduction of experimental rate constants in reaction studies,  and provide immediate pointers to follow-up measurements that are needed.

We start with the perspective that there is a one-to-one correspondence between the image observed and a set of input parameters - the parameters describing the physical geometry and fields applied by the trap, and the number of ions of different masses in the crystal.  
Although that relationship is effectively determined by Newtonian laws of classical physics, we seek to train a Neural network to recognise  the input parameters - and particularly the number of ions - simply by `looking at' the images, obviating the need to carry out any molecular dynamics simulations for the analysis of experimental images. 
 The inference of ion numbers is immediate - it takes a matter of seconds for each image on a standard laptop computer - and effectively gives ion numbers in `real time.' Furthermore, virtually all the computational cost is associated with the CNN training process, and, once this is completed, the application to infer ion numbers comes at a negligible cost.

Our initial objective was to train a CNN with a sufficiently broad training set (with images generated using a wide range of ion trap parameters), so that it be can used across multiple ion-trap set-ups and can determine ion numbers from images without  being provided with any further information about the trapping parameters used to obtain those images.  This is analogous to the use of CNNs to identify everyday objects or animals from a photo without being given prior information about the details of the specific camera used, or how close the object was to the camera, or what the resolution of the camera was.
\subsection{Neural networks and their applications}\label{neural-networks}

The use of neural networks and artificial intelligence to solve demanding problems has been ubiquitous over the last decade, but the idea of using the human-brain model for a system that can be trained to provide the desired output for 
a given set of inputs can be traced back to 1943 \cite{McCulloch1943}.  
The first trainable `neural network' with hidden layers was constructed in the 1980s \cite{Rumelhart1986}.
Neural networks are a type of supervised learning technique, and require an extensive training-data set,  consisting of input data points and their associated outputs.
In the Coulomb crystal image case, the inputs are the CC images in numerical matrix form, and the output data points are parameters such as the ion numbers, or as shown recently by Yin and Willitsch, the secular temperature of the crystal \cite{Yin2025}. 

Training takes place by presenting a set of  input data points and known associated outputs  to the network,  which attempts to either classify the input data (for example identifying whether a photographic image shows a cat or a dog) or to produce  values for the output parameters (regression), and this is compared to the known output values and loss is calculated. 
The `loss' quantifies the difference between predicted values generated by the CNN model,and the actual values. It can be defined as either the cross-entropy in the classification case or the Root Mean Square Error (RMSE) in the regression case. 
The numerical weights in the network are then adjusted to minimise this loss using the first-order derivative, and the process is repeated. The amount that these weights are altered (with respect to the gradient of the loss) is known as the learning rate.

By 1998, LeCun et al\cite{Lecun1998} had developed a system based on the concept of receptive fields, which are used by the human eye to identify features in an image, to make progress in a handwriting analysis problem. These systems became known as Convolutional Neural Networks(CNNs).
Layers of CNNs analyse the data on different length scales before the analysis propagates into the traditional dense layers of the network prior to an output classification or output parameter values being produced.  A turning point for the practical application of CNNs came when a huge amount of data started to be created by digital cameras, especially in phones, which was easily collated on the internet. This was coupled with the realisation that even consumer Graphical Processing Units (GPUs) could substantially accelerate the training process by orders of magnitude of using CPUs. The massive amount of training data and the ability to quickly and affordably train networks, meant that the architecture of these networks could increase in complexity. 
CNNs became the leader in this field after the success of AlexNet in the 2010 ImageNet competition where, across a 1.2 million image sample comprising of 1000 classes, it demonstrated an accuracy that was significantly better than the state of the art \cite{Krizhevsky2012}.
This marked the transition from CNNs being an interesting idea into a practical technique, and substantially reduced the computational cost of image recognition.

The objective of the work described in this paper is to build a suitable neural network, or redeploy a pre-existing neural network, that can be trained on simulated and/or experimental CC images with known ion numbers and trap/imaging parameters for each image.  Having trained the network, we seek to use it to predict the ion numbers for unseen images, and without knowledge of other trapping parameters.  Ideally, we aim for this analysis to  be achievable in real time while the experiments are taking place - which, in practice, means analysis in a few seconds per image.

\section{Computational and experimental methods}
\subsection{Molecular dynamics simulation and training of the convolutional neural network}
To solve the problem outlined, we started by constructing a very large dataset of simulated images using the Coulomb Crystal Molecular Dynamics code (CCMD), described most recently in Reference \citen{Oldham2014},  on the University of Birmingham's BlueBear High-Perfomance Computing (HPC) facility.
The CCMD code simulates the laser cooling of a specified number of ions (typically 100-600 ions in this work), using a set of specified input parameters characterising the trap dimensions and fields, the imaging parameters, the masses of the ions, the forces due to Coulomb repulsion between ions, laser cooling, and stochastic heating processes (which  have their origins in primarily collisions with background gas). 
The code is a classical molecular dynamics routine, rewritten by M. Bell and C. R. Rennick (University of Oxford) in C$^{++}$, and is based on a pre-existing MD code Protomol \cite{Matthey2004}.
The force is written as 
\begin{equation}
    \mathbf{F_{tot}}
=\mathbf{F_{trap}} + \mathbf{F_{ion}} + \mathbf{F_{cool}} + \mathbf{F_{heat}}
\end{equation}
where the terms in the equation represent the trapping forces, the Coulomb repulsion, the laser cooling force and the stochastic heating force respectively.
The equations of motion are integrated using a velocity-Verlet algorithm across the time cycle of laser cooling and then across a subsequent cycle of image gathering.  Typically, each cycle involves around one million time-integration points of 0.2~ns each.

The positions and velocities of the particles, sampled over the image-gathering period,
are then used to produce a simulated image. The image shows the 2D-sliced probability density  of the
ions,  with each slice corresponding to ions at a specific distance from the
focal plane, nominally the centre of the crystal. A Gaussian blur is applied to sliced
images that are not in the focal plane, with greater blurring for slices that are further
from the plane. These slices are then summed together to give a simulated image to
compare with experimentally obtained images.

 We mainly simulate images for  pure Ca$^+$ crystals  in this paper, although the CCMD code is capable of simulating images of mixed crystals with a combination of laser-cooled ions and sympathetically cooled ions (such as in \Cref{fig:CC_intro}(c)), and we made use of that capability as demonstrated later.

For the CNN training set, we chose ranges for the key input parameters that may vary in a range of typical experimental Paul-trap CC studies.
The parameters initially varied were as follows;

\begin{itemize}
    \item 33 ion number values ranging from 30 to 350 in steps of 10.
    \item 14 radio-frequency peak-to-peak voltages ranging from 170~V to  300~V in steps of 10~V.
    \item 21 endcap voltages, ranging from 1.5~V to 3.5~V in steps of 0.1~V.
    \item 24 eta parameters (aparameter that is needed to characterise  the geometric imperfection of experimental images) ranging from 0.240 to 0.355 in steps of 0.005~V.
    \item 3 random seeds, used to initiate randomly the velocities and positions of the ions in the simulations.  
\end{itemize}

Taking all permutations of these values leads to a data set of 698,544 images. 
In order to test the quality  of the training process, the data set is partitioned in a 5:1:1 ratio into a training set,  a validation set, and a test set, where the `validation' set is  used as unseen data to test the accuracy of the predictions of the neural network,
and the `test' set is reserved for further testing of the hyperparameters of the model if required. These sets were defined by including each of the parameters in the filename of each image, taking the hash of that filename, and taking modulus 7 of that value. The validation and test sets had a modulus of 0 and 1, and this approach had three advantages: it is a simple method; we could start validation before the training simulation had completed without risking leakage between the two data sets; and, most importantly, this should guarantee that there is no bias between the training, validation and test samples.  RayTune was used to carry out an unbiased optimisation of hyperparameters \cite{liaw2018}.

Before training, the simulated images are automatically cropped with a rectangular boundary around the elliptical image in order to reduce the number of black (zero) pixels analysed by the neural network. We use image augmentation to diversify the  data that the network is trained on and to minimise any possibility of overtraining, even with the large number of training images.  This includes rotating the crystal image, moving it off centre, and enlarging or reducing its size.  This augmentation is applied randomly before adding a simulated image to the data set. 

Initially, we attempted to build our own CNN, which was coded using PyTorch or PyTorch Lightning. Typically, the models we employed included three CNN layers feeding into three `standard' dense layers performing a regressional analysis to produce a single output (the ion number). 
While these relatively simple models performed moderately well, especially in the ion-number analysis of a validation set of simulated images, they did not perform satisfactorily when given experimental images to analyse.  

Subsequently, it was decided to test out the RESNET50 CNN model \cite{Nicolas2024}, and the remarkable capabilities of that model became quickly apparent. This is the model that was used for all the results presented in this paper.
The RESNET50 model was developed by Microsoft Research in 2015 and is a deep CNN with  the number ``50”  referring to the number of layers in the network. The model, which has been applied to diverse applications such as facial recognition and complex-scene analysis,  has been trained on the ImageNet dataset that now contains over 15 million images and 1000 classes (images of objects, animals, etc. )\cite{Krizhevsky2012}.
Only a modification of the input and output layers was required in the present work to use the RESNET50 model for our purposes. For the output layer, we replaced the final classification step, with a conventional dense numerical output layer, and the predicted ion number was output as a double-precision number. The top layer was changed to work with greyscale images, as well as colour images.
As shown below, this model has been successfully deployed to analyse data from laboratories in Liverpool/Oxford, UK and Boulder, CO USA.  A self-standing  `Inference' routine has been created in this work, which runs on a standard laptop in any location, via a link using the Docker platform to the code. It uses the weights generated in the adapted RESNET50 model through the training on the simulated CC images to determine the Ca$^+$ ion number for any CC ion image presented to the routine. The time taken to analyse each image is typically less than five seconds.

\subsection{Generating a set of experimental CC images for verification.}\label{subsec:experimental}

An additional set of experimental images was generated  to test the accuracy of ion-number prediction by the CNN on real data.  These were recorded on the laser-cooled ion-trap time-of-flight mass-spectrometer set-up at JILA, University of Colorado Boulder described in detail in reference \citen{Schmid2017}. 
Ca$^+$ ions formed via non-resonant ionization of an effusive Ca beam using the third harmonic of a Nd:YAG laser were loaded into a linear Paul trap. The Ca$^+$ ions were then laser cooled to secular temperatures of $<$~1 K using the frequency-doubled and fundamental outputs of two Ti-Sapphire lasers ($\sim$4~mW at 397~nm and $\sim$20~mW at 866~nm). The resultant Coulomb crystal typically consisted of several hundred ions. The fluorescence emitted by the ions of the Ca$^+$ Coulomb crystal was imaged onto an electron-multiplying charge-coupled device (EMCCD).  The number of ions in each  crystal was determined  via time-of-flight mass spectrometry after the image had been recorded.  

\begin{figure}[hbt!]
    \centering
    \includegraphics[trim=20 80 30 90,clip]{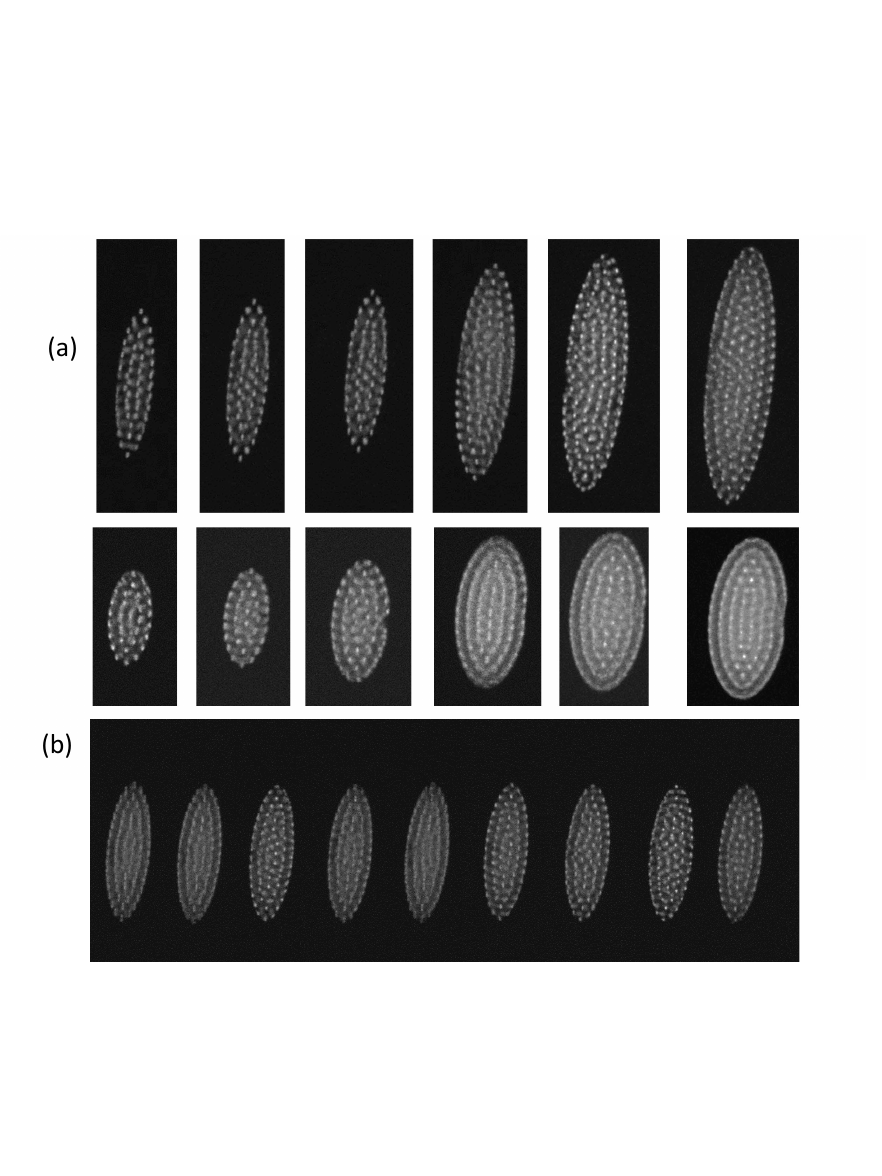}
    \caption{(a) Experimental images recorded in Boulder for pure Ca$^+$ crystals with various ion numbers, obtained  with an RF peak-to-peak voltage of 400~V (upper line of images) and 250~V (lower  images). (b) A sequence of Boulder experimental images recorded for a single Coulomb Crystal at 2~s intervals.}
    \label{fig:Boulder_images}
\end{figure}

A set of 12000 images was generated,  involving creation of 200 individual pure Ca$^+$ crystals of a range of sizes from {\it{ca}} 50 to 900 ions. These consisted of 10 sets of 20 crystals with a combination of variable endcap voltages [1.75~V, 2.5~V, 3.25~V, 4.0~V and 5~V] and radiofrequency peak-to-peak voltages  [400~V and 250~V]. The size (ion number) of each crystal was set approximately by varying the length of time for which the Ca atom beam was ionized. A set of 60 images was recorded
for each crystal at 2 second intervals (total observation time 2 minutes per crystal) and the ions were then ejected into a time-of-flight mass-spectrometer and a calibrated Ca$^+$ ion number obtained.  To confirm this
calibration, the method of iterative simulated-image matching was used to compare with the ion numbers derived from the mass spectrum. A sample of some of the images for different crystals obtained at different RF and endcap voltages and ion numbers  is shown in  \Cref{fig:Boulder_images} (a).  A time sequence of images at 2 second intervals is also shown for one crystal in  \Cref{fig:Boulder_images} (b).

\section{Results}
In this section, we present the results of applying the Inference routine to determining Ca$^+$ ion numbers in five different sets of CC images. We begin with the application to the verification set of 100,000 simulated images of pure Ca$^+$ crystals, and then test the level of success the CNN has with simulated mixed-species crystal images (Ca$^+$ with sympathetically cooled ND$_3^+$ or CaF$^+$). 
It should be noted that the training set included only pure Ca$^+$ crystals. We then apply Inference to {\it experimental} crystal images, first those recorded in Boulder, and then those recorded in Liverpool/Oxford, and noting that the CNN was  trained on primarily simulated images with the Liverpool/Oxford trap parameters. Finally, we apply it to experimental mixed-species crystals from Liverpool/Oxford with up to 20\% of Kr$^+$  doped into the Ca$^+$ crystal. 
These data sets are presented in a progression of what we expected would be an increasing challenge for the CNN.

\subsection{Ion-number inference with simulated pure calcium Coulomb crystals}
\begin{figure}[hbt!]
    \centering
\includegraphics[trim=35 130 50 220,clip]{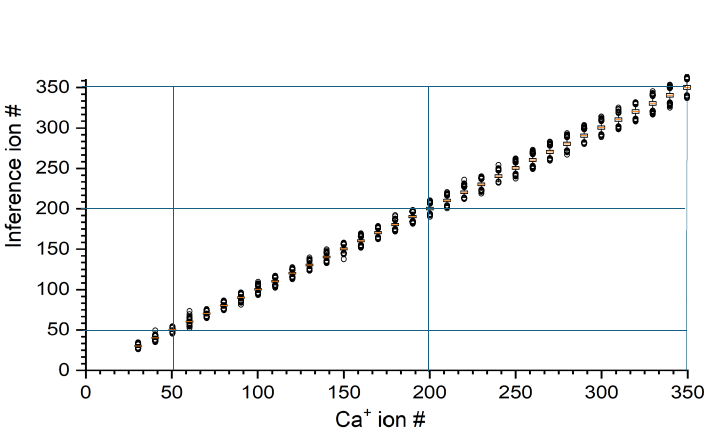}
    \caption{Plot of inferred  Ca$^+$ ion numbers for 100,000 simulated images (from validation routine) versus the Ca$^+$ ion number used to simulate each image.  The centres of the pink rectangles show the mean inferred ion numbers for the ca 3000 images in each bin. The vertical thickness of the rectangle shows the  standard deviation for ca 3000 images.  Black open circles show the inferred ion number for the 10 greatest outliers in each bin above and below the mean.}
    \label{fig:validation_plot_sims}
\end{figure}

\begin{figure}[hbt!]
    \centering
    \includegraphics[trim=20 20 30 80,clip]{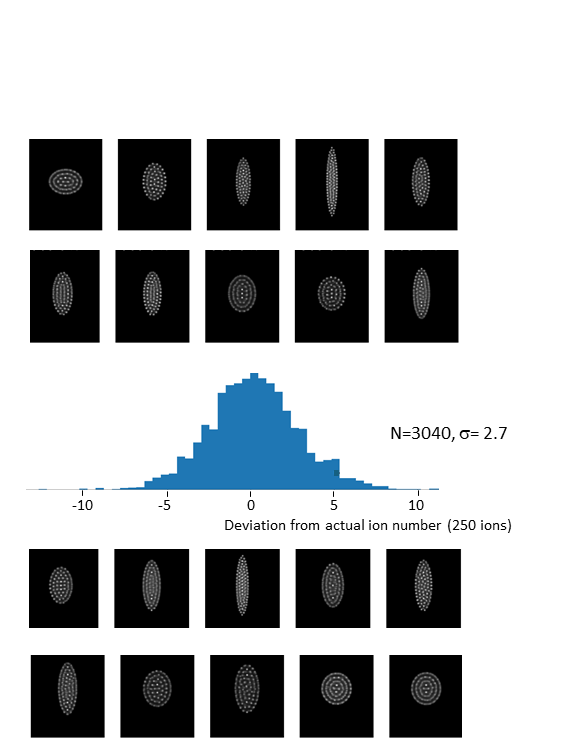}
    \caption{Histogram of output ion numbers from the validation routine for 3040 simulated images with 250 ions. The images for which the deduced ion number has the greatest positive and negative variation from the actual ion number are shown above and below the histogram.}    \label{fig:Inference_outliers}
\end{figure}

The verification-set of {\it ca} 100,000 simulated images was run through the CNN verification code to determine the Ca$^+$ ion numbers and  compare them with the ion number that was used to generate them.  
The images were binned into sets  according to the actual ion number, with approximately 3000 images in each of 33 bins in step sizes of 10 ions (30~\textendash~350 ions). Within each bin, the images had widely varying trap parameters as described earlier.  The standard deviation of the CNN-inferred ion numbers from the actual ion numbers was calculated for each bin, and these are plotted as the vertical thickness of the pink rectangles in \Cref{fig:validation_plot_sims}. These deviations are in the range 0.9~\textendash~3.7 ion-number units, or a percentage deviation of 1.2~\textendash~1.8\%. Example values of the standard deviations in ion  numbers are: 0.9 at total ion number 50, 2.0 at 150, 2.7 at 250, and 3.7 at 350.    Each pink bar is centred on the average value of the ion number produced by the Inference routine and this
follows a straight line of gradient 1 and passes through the origin. 
The 10 most outlying points above and below the mean are also shown for each bin in the figure with black circles.
A sample of the crystal images  represented by the outlying error points is shown in \Cref{fig:Inference_outliers}. 
Although these outlying crystal images  are typically almost circular in geometry (spherical crystal) or prolate-shaped (long and thin, pointing vertical), it is not obvious why such categories of image should show higher errors, except perhaps that the RF and endcap voltages used to generate these lie at the extremes of the range incorporated in the training set. In summary, the CNN is able to predict ion numbers of unseen simulated images, generated with a wide spread of trap parameters, with a percentage error of  1-2\% in a few seconds per image.
\subsection{Analysis of simulated mixed-species Coulomb crystals (Ca$^+/$CaF$^+$ and Ca$^+/$ND$_3^+$) }
An intriguing question is whether the neural network can determine calcium-ion numbers in images of mixed-chemical-identity crystals. As explained earlier, ions of higher mass than calcium - e.g. calcium fluoride CaF$^+$, mass 59 u, tend to take positions in layers around the edge of the crystal, and the fluorescing calcium ion framework takes on the appearance of a flattened ellipsoid. With this very different shape from the pure calcium crystal, it might be expected that a CNN that has not been trained to recognise such crystals would have difficulty in generating physically reasonable answers for the calcium-ion numbers.

\begin{figure}[hbt!]
    \centering    
    \includegraphics[trim=30 30 30 50,clip]{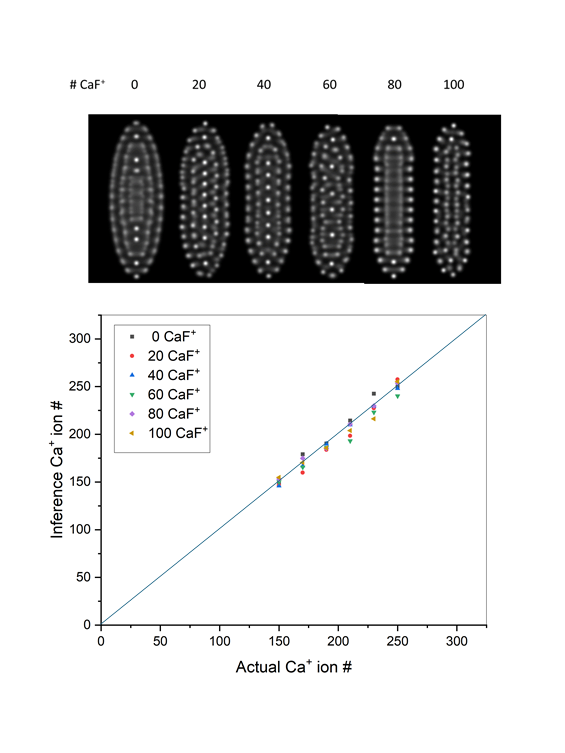}
    \caption{(a) Simulated crystals of Ca$^+$ which contain a proportion of sympathetically cooled CaF$^+$. Each crystal has 250 ions in total and CaF$^+$ ion numbers ranging from 0~\textendash~100 as indicated. (b) Inference results, showing deduced ion number versus actual Ca$^+$ ion number for simulated  Ca$^+$/CaF$^+$ crystals with combinations of 0~\textendash~100 CaF$^+$  ions and  150~\textendash~250 Ca$^+$ ions. }
\label{fig:simulated_mix_images_CaF}
\end{figure}

A test set of  simulated mixed-crystal images was generated with combinations of 150 to 250 Ca$^+$ ions and 0 to 100 CaF$^+$ ions; a sample of these are shown in \Cref{fig:simulated_mix_images_CaF} (a).
The application of the Inference routine to deduce the Ca$^+$ numbers for this test set of images is shown in \Cref{fig:simulated_mix_images_CaF} (b).
It is clear that even for this small number of images tested (36), there is a reasonably good prediction of Ca$^+$ ion numbers with an
RMS error of 6.4 and an RMS percentage error of 3.1\%. 

\begin{figure}[hbt!]
    \centering    
    \includegraphics[trim=30 30 30 40,clip]{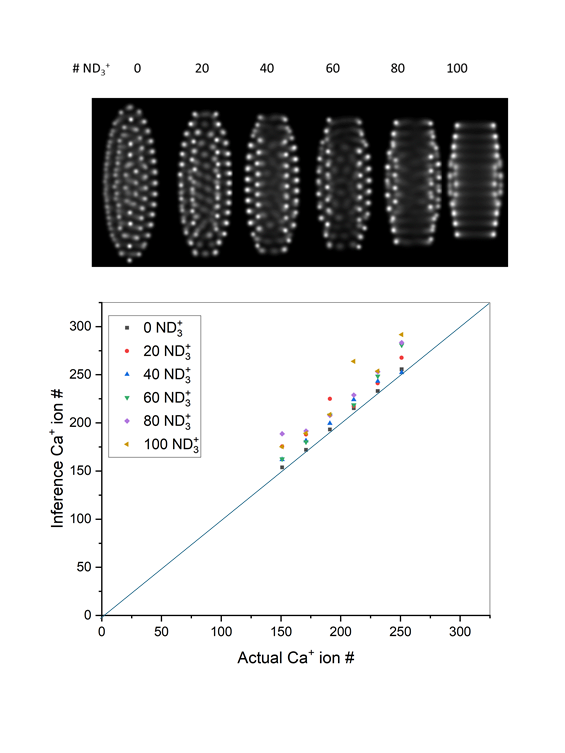}
    \caption{(a) Simulated crystals of Ca$^+$ which contain a proportion of sympathetically cooled ND$_3^+$. Each crystal has 250 ions in total and ND$_3^+$ ion numbers ranging from 0~\textendash~100 as indicated. (b) Inference results, showing deduced ion number versus actual Ca$^+$ ion number for simulated  Ca$^+$/ND$_3^+$ crystals with 0~\textendash~100 ND$_3^+$  ions and  150~\textendash~250 Ca$^+$ ions.  }
    \label{fig:simulated_mix_images_ND3}
\end{figure}

In the case of mixed crystals with low-mass sympathetically cooled ions, the images give the appearance of having a hollow core. This is illustrated for a set of simulated images for Ca$^+$ with ND$_3^+$ (mass 20 u) shown in \Cref{fig:simulated_mix_images_ND3}(a). 
Given that the neural network had not been trained on such hollow-core crystals, we expected that the prediction of calcium-ion numbers from such images might be more difficult than for the flattened images in \Cref{fig:simulated_mix_images_CaF}.   \Cref{fig:simulated_mix_images_ND3} (b) shows the outcome of predicted ion numbers for 36 crystals of Ca$^+$/ND$_3^+$ composition with combinations of 150 to 250 Ca$^+$ ions and 0 to 100 ND$_3^+$ ions. The RMS ion number error is 22 for this set, and the RMS percentage error is 11\%. 
While the accuracy of prediction is not as good as for the set in \Cref{fig:simulated_mix_images_CaF} (b), we consider the accuracy to be  reasonable, as is the ability to get the correct ordering of the Ca$^+$ ion numbers in general. Furthermore the gradient of the trend is still approximately 1, and only the intercept is shifted.
The tendency for the model to over-estimate the ion numbers is perhaps not surprising given that the size of the ellipse encompassing the ions is somewhat larger than the CNN would `expect' for a crystal with that number of Ca$^+$ ions.  
In principle, if this overestimation could be demonstrated to be systematic, then it could be corrected, enabling use of the CNN for identifying numbers in such mixed crystals with light sympathetically cooled ions. As discussed later, the CNN could also be retrained to include a set of simulated mixed crystal images in the training set.

\subsection{Analysis of experimental pure Ca$^+$ Coulomb crystals images recorded in Boulder}

A test set of 370 experimental images, recorded as described in \Cref{subsec:experimental}, was selected from the 12000 images available.  The set comprised the first 10 images (out of 60) recorded at two-second intervals for  35 separate crystals.  
We chose to focus  on the images recorded with 250~V RF peak-to-peak voltage, because the 400~V RF  peak-to-peak voltage value was beyond the range of parameters used in our training set, and it also generated a number of long, thin, vertically aligned crystals that pushed towards, or over, the boundaries of the imaging system. From each set of 20 crystals, we  selected 6~\textendash~10 crystals from each set with a given endcap voltage (1.75~V, 2.5~V, 3.25~V, 4.0~V, 5.0~V) that covered the range from the smallest crystal size to an ion number of approximately 450 at roughly intervals of 50. It should be noted that it is not possible experimentally to prepare a crystal with a precise number of ions on demand.

Prior to running through the Inference routine, 
the experimental images are converted so that they have the same magnification and pixel array dimensions as used in the training set (which was based on the pixel array dimensions for Liverpool images).
Crystals were rotated by ca 5 degrees to bring the ellipse axis into the  vertical orientation, and any non-zero pixels outside the ellipse of the crystal are set to zero. A five-point two-dimensional median averaging was also applied to reduce the noise level.

\begin{figure}[hbt!]
    \centering
\includegraphics[trim=20 28 30 50,clip]{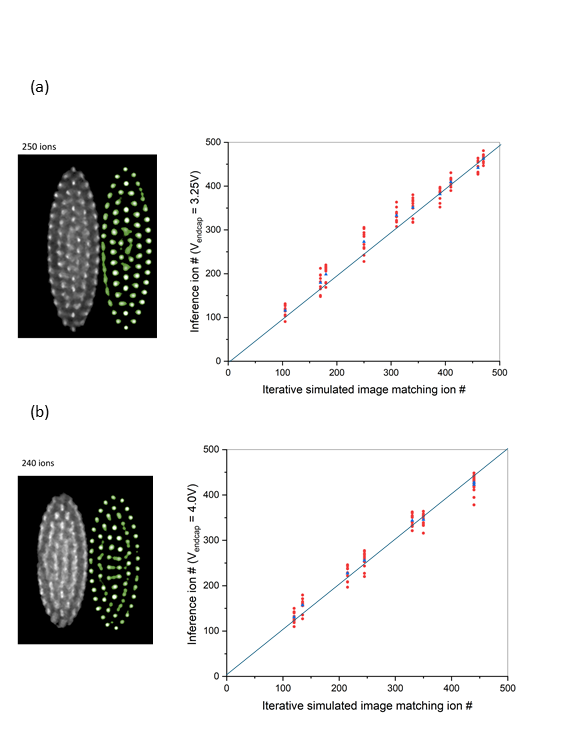}
    \caption{Results for Ca$^+ $ crystals recorded in Boulder, showing deduced ion number from Inference versus Ca$^+$ ion number derived by the method of iterative simulated-image matching (see insets for comparison of example simulated image (green) with observed image (BW)). (a) The results for 10 images recorded at 2 second intervals are shown for each of 10 crystals. 
     The ion trap parameters were RF peak-to-peak voltage 250~V, endcap voltage 3.25~V. A line of gradient 1 passing through the origin is shown. (b) as in (a) but with 7 crystals recorded at endcap voltage 4.0~V.}
    \label{fig:Inference_for_Boulder}
\end{figure}

\Cref{fig:Inference_for_Boulder} (a) and (b)  show examples of the plot of CNN-predicted ion number, for crystals obtained using a common endcap voltage (3.25~V and 4.0~V respectively),  against the ion numbers determined by the method of iterative simulated-image matching (an example of the outcome of such matching is shown in the inset of each figure).  We have chosen to use the iterative simulated-image matching ion numbers for horizontal axis in these plots, rather than the mass-spectrometer numbers, to enable a fair comparison with results for the Liverpool/Oxford data (where we do not have calibrated mass-spectrometric ion numbers currently.)
\Cref{tab:Inference_for_Boulder} lists the RMS- and mean-percentage errors for all five data sets. It can be seen that the RMS error is typically around 10\%.  
A significant part of that error stems from the dispersion of values determined for the 10 images of each crystal - as is apparent from those figures.   If the mean ion-number prediction is used for those 10 images in place of the individual values for the 3.5~V set,  the mean and RMS percentage errors reduce to 5.3\% and 6.4\% respectively.  If it were possible to identify in advance which of the 10 images in each of these sets would give the best predicted ion numbers then the errors could be reduced to around 1\%. 
However, inspection of the crystal images does not give any obvious grounds for making that selection.
Nevertheless, given that the CNN has not used any experimental images in its training, and when considering  all the
imperfections of such experimental data, we believe this is an encouraging result. It must also be recognised that there is an `error', of up to 5\%, in the 
ion numbers determined by the iterative  simulated-image matching method, and these are assumed here to be the `correct' numbers for calculating the errors.  In practice, an exact match of simulations and experimental images is rarely achievable, given that there are inevitably many near-degenerate structures for the crystal at the crystal sizes and temperatures used here (and the CCMD simulation will land on just one of these), and also the imperfections of experimental images  play a role in contributing to the error.

\begin{center}
\begin{table}[ht]
    \centering
    \begin{tabular}{c|c|c|c|c}
        End cap voltage /V & Number of images & Ion number range & Mean \% error&RMS \% error \\
        \hline
        \hline\\
        1.75 & $8\times 10$&$110-411$& 11.6& 15.3\\
        2.5 & $6\times 10$&$115-450$& 9.2& 12.3\\
        3.25 & $10\times 10$&$105-470$& 7.3& 9.9\\
        4.0 & $7\times 10$&$120-440$& 7.5& 10.2\\
        5.0 & $6\times 10$&$115-392$& 7.2& 8.9\\
       \end{tabular}
    \caption{Mean and RMS errors for Inference predictions of Ca$^+$ ion numbers in five sets of images recorded in Boulder}
    \label{tab:Inference_for_Boulder}
\end{table}
\end{center}

\subsection {Analysis of pure Ca$^+$ crystals recorded in Liverpool/Oxford}

\begin{figure}[hbt!]
    \centering    
    \includegraphics[trim=20 30 30 60,clip]{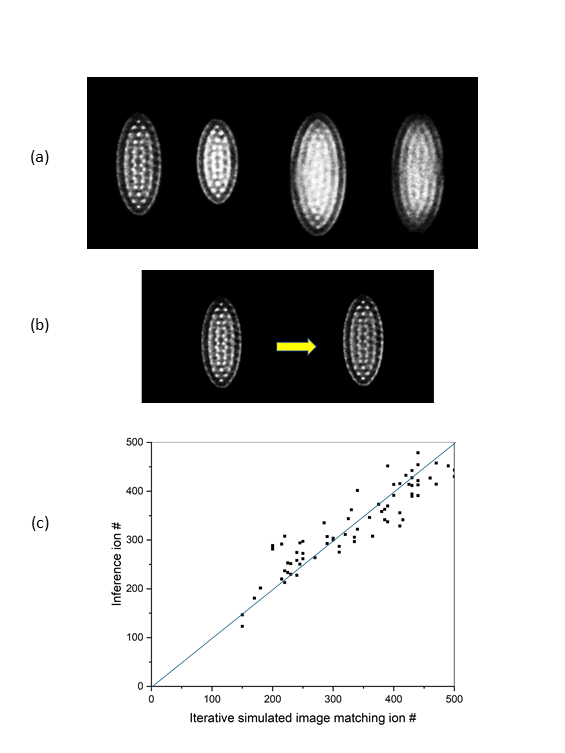}
    \caption{(a) Example experimental images of Ca$^+$ crystals from Liverpool/Oxford, recorded in the period 2019~\textendash~2022 (unpublished). (b) Example of the effect of the pre-manipulation of experimental images from Liverpool/Oxford as described in the main text. (c) Inference results for pre-manipulated Ca$^+ $ crystal experimental images recorded in Liverpool/Oxford, showing deduced ion number from Inference versus Ca$^+$ ion number derived by iterative simulated-image matching.  A line of gradient 1 passing through the origin is shown.}
        \label{fig:Liv_experiment}
     \end{figure}

To test the transferability of our trained model across experimental setups and trap parameters, the Inference routine has also been applied to deduce ion numbers for a set of 90 images that were recorded (see References \citen{petralia2020,tsikritea2021} for further details)  on an ion-trap set-up in the Heazlewood Group in Oxford or Liverpool during the period 2019~\textendash~2022, on 18 separate days. Compared to the Boulder experiments, the physical dimensions of the Liverpool ion trap are slightly different, the cooling laser power is lower, and a different camera is used with lower resolution. 
For each of the images, the ion number had been previously deduced by the method of iterative simulated-image matching.
The trap parameters used to record these images were a radiofrequency peak-to-peak voltage  $V_{RF} = 200-210$~V, and  endcap voltage $V_{end} = 2.25-2.95$V 
and deduced ion numbers ranged from 150~\textendash~720 ions.     
As illustrated by the sample of images in \Cref{fig:Liv_experiment} (a), the sharpness of the images and signal-to-noise level is more variable than the images recorded in Boulder, principally as a consequence of the older imaging camera deployed, and there is a tendency
for the outer elliptical layer to appear less intense than we would expect from simulations.
%
In order to increase the similarity of the images to the simulated images on which the CNN had been trained,  we performed some additional pre-processing steps compared to the Boulder-recorded images.
In particular, the intensity around the outer layer was artificially enhanced to improve uniformity of intensity using a square function of the ellipse equation ($f(x,y)$ which is equal to 1 at the ellipse boundary):
\begin{equation}
I(x,y) = I_0(x,y) + A\times I_0(x,y) \times (f(x,y))^2
\end{equation}
where
\begin{equation}
f(x,y)=\frac{(x-h)^2}{a^2} + \frac {(y-k)^2}{b^2}
\end{equation}
and the ellipse is centred at co-ordinates $(h,k)$ and has major and minor axes of length $2a$ and $2b$. $I_0(x,y)$ is the un-manipulated intensity and A is a constant that had an identical magnitude for all images.
In addition, the intensity of images is normalised, such that the ‘brightness density’ = sum of non-zero pixels/number of non-zero pixels is constant across the image set.
The overall effect of these changes is illustrated for one crystal in \Cref{fig:Liv_experiment} (b).

\Cref{fig:Liv_experiment} (c) shows a plot of deduced Ca$^+$ ion number from  the Inference routine, versus that deduced previously by the method of iterative simulated-image matching.  
Given that the images are significantly more variable in quality than for the Boulder set, this level of accuracy and the matching of the general trend of the data to a straight line of gradient 1 passing through the origin is, in our view, satisfying.
The mean percentage error in ion-number prediction for this data set is 10.4\%, which is comparable with that achieved for the Boulder data. 
\subsection{Analysis of mixed Ca$^+$/Kr$^+$ crystals recorded in Liverpool/Oxford.}
\begin{figure}[hbt!]
    \centering    
\includegraphics[trim=20 30 30 60,clip]{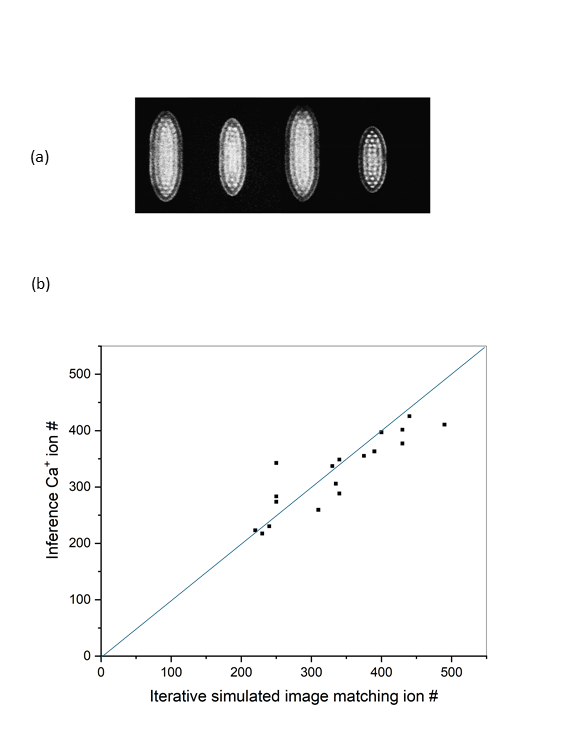}
    \caption{(a) Example experimental images of mixed-species  Ca$^+/{\rm Kr}^+$ crystals from Liverpool/Oxford, recorded as described in References 
    \cite{tsikritea2021,tsikritea2022} in the period 2019~\textendash~2022.  (b) Results for pre-manipulated Ca$^+$/Kr$^+$  crystal experimental images recorded in Liverpool/Oxford, showing deduced ion number from Inference versus Ca$^+$ ion number derived by iterative simulated-image matching. The crystals have varying numbers of Kr$^+$ ions that are less than 20\% of the total ion number. A line of gradient 1 passing through the origin is shown.}
    \label{fig:Liv_experiment_mix}
\end{figure}

Finally, we also tried applying the Inference routine to a set of experimental mixed-crystal images, analogous to those shown in \Cref{fig:simulated_mix_images_CaF}.
\Cref{fig:Liv_experiment_mix} (a) shows four examples of the crystal images that were used. 

These were obtained in a series of experiments conducted in Oxford in 2020. The selected images were all mixed crystals of  Ca$^+$/Kr$^+$. We eliminated images from the sample that contained greater than 500 calcium ions, those that contained more than 20\% rare-gas ions,  and those that appeared visually to be highly asymmetric in the vertical direction.

\Cref{fig:Liv_experiment_mix} (b) shows a plot of Ca$^+$ ion number deduced by the CNN Inference routine for mixed crystals of Ca$^+$/Kr$^+$ versus the ion number deduced by iterative simulated-image matching.  The RMS-percentage deviation of Ca$^+$ ion number is 12.5\% and the mean percentage deviation is 9.3\%. We are not able to deduce the rare-gas ion numbers using the CNN, because it has not been trained to do that.  In principle, it would be possible to retrain the CNN with mixed crystals, and to train it to deduce both the Ca$^+$ ion number and the dopant (e.g., rare-gas ion) number.    However, we have not performed that training to date.

Nevertheless, this limited exercise demonstrates once again the  versatility of the model that can be extended to make ion-number predictions for experimentally-imperfect  mixed crystal images (Ca$^+$ / Kr$^+$) of the type it had not seen in training.

\section {conclusion and future perspective}
Our overall objective at the outset of this project was to develop (or redeploy) a global CNN model that was capable of determining ion numbers from Coulomb Crystal images in real time during experimental recordings.   
The goal was to train the neural network on a sufficiently diverse dataset, enabling it to be used for images obtained in different laboratories without retraining. This would ensure adaptability to variations in quadrupole ion trap geometries, applied fields, and, ultimately, mixed-species ion crystals.
The work reported in this paper demonstrates that this goal is a challenging, but potentially achievable, objective. We have successfully used a slightly modified CNN model (RESNET50) that has been retrained on a very large set of simulated images, which were  generated primarily using the trap parameters of the Liverpool/Oxford laboratory,  and applied it with some success to experimental results from both that laboratory and from Boulder. This was achieved without inclusion of  experimental data in the training set. The typical errors of {\it ca} 10\% are larger than estimated standard deviations from 
mass-spectrometry or the method of iterative simulated-image matching ({\it ca} 5\%). But there are some key potential advantages in the CNN approach, including the non-destructive nature of the ion-number determination (in contrast to the  mass-spectrometric approach), and the fast analysis of images - in seconds, not days - in contrast to  the iterative simulated-imaging matching. 
Furthermore, we have shown that, even though we  trained the CNN  only on simulated images of single-component crystals (pure Ca$^+$ crystals), it could be applied in limited circumstances to identify the Ca$^+$ ion numbers for two-component crystals, especially those where the second component was a species heavier than Ca$^+$, for both simulated and experimental images.

Although this is a very promising outcome, the challenges of generating and working with a training set of more than half a million images should not be underestimated. The training set took many months to generate on the University of Birmingham's high-performance computing set-up (BlueBear). Ideally, we would follow up the present work by generating an extended training set on which to retrain the model, with inclusion of simulated images of multi-component crystals to add to the existing training set, to improve the accuracy and versatility of identifying ion numbers in such crystals. We would also generate enough experimental images, with pre-determined ion numbers (via another method),  to add these to the training set.  However, we would probably need to have at least 10\% of the total images in the training set being experimentally generated to have any impact on the training (i.e.  50,000~\textendash~100,000 such images). Such large sets of ion-number-labeled images do not exist currently. 

However, even with current levels of accuracy, the approach described here could be used in a hybrid
methodology for identifying the ion numbers,  in parallel with other measurements. 
In the case of mass-spectrometric analysis, the CNN-based evaluation of ion numbers would be extremely useful for routine checking of the calibration of calcium-ion peak heights as a function of crystal size.
In the case of the iterative simulated-imaging, the CNN inference ion number, could provide an excellent starting point for further refinement through image matching even for mixed-species crystals. We also note that a typical sequence in chemical kinetic measurements, as shown in \Cref{fig:CC_intro} (c), always starts with a pure calcium crystal, and knowing the initial Ca$^+$ number can streamline the process of image simulation of the mixed crystals.
Furthermore, the very large database of 700,000 simulated Ca$^+$ crystals forms a valuable, diverse library of images, which could be used to circumvent the need to simulate new images for image matching within the parameter ranges covered in the database. 

In other complementary work conducted in Basel in parallel with this work, a more targeted CNN approach was adopted \cite{Yin2025}. A smaller training set (7200 images) was created, consisting of simulated CC images generated with a relatively narrow set of ion-trap parameters, and using the RESNET 18 and ALEXnet  models. Ca$^+$ ion numbers and secular crystal temperatures were deduced from simulated and experimental images of pure Ca$^+$ crystals. When applied to a validation set of simulated Ca$^+$ images with ion number 100~\textendash~299 and fixed ion trap parameters, the RESNET 18 model inferred the correct ion number for 93\% of the tested images (classified in steps of one ion-number unit) .
In some circumstances, this targeted approach may be advantageous where a fixed set of ion trap parameters is used extensively and repeatably in a particular experimental ion trap. However, the use of a targeted training regime implies that if the ion-trap parameters are varied significantly outside that narrow range (as  they frequently are in the Boulder set-up for example) then a completely new training set would most likely need to be generated each time, and the model cannot easily be  transferred across different experimental set-ups without retraining and verification.

Finally, we observe that the model that was eventually used in the current work - the RESNET50 model - was not originally developed for image recognition of scientific images, but rather for recognition/characterisation of images of everyday objects, animals etc, facial recognition, and complex-scene analysis. The reason why it works also for these scientific images is that, in both cases,  the different layers of the CNN are used to characterise features of the image at different length scales, before compiling the information to achieve an overall recognition. This multi-lengthscale characterisation is needed in the process to recognise a Coulomb crystal image with a given ion number, as much as it is to recognise a human face. 
It seems likely that this could be a versatile tool for a wide range of other imaging applications in the physical sciences, as has already been demonstrated in the area of medical imaging. \cite{Xua2023}

\section*{Acknowledgements}
T.P.S is grateful to the Leverhulme Trust  for a grant (EM-2022-027) to support this work,  for a Visiting Fellowship at JILA, University of Colorado, Boulder in 2023, and for the expertise and support of the University of Birmingham Research Software Engineering Group. B.R.H. is grateful to the Engineering and Physical Sciences Research Council (EPSRC project EP/N032950/2), the European Commission (ERC Starting Grant project 948373), and the Leverhulme Trust  (projects RPG-2022-265 and PLP-2022-215) for funding.

\clearpage
\bibliography{references} 

\end{document}